\pdfoutput=1

\documentclass[preprint,12pt, a4paper]{elsarticle}

\usepackage{amssymb}
\usepackage{hyperref}

\usepackage{url}
\usepackage[T1]{fontenc}
\usepackage{amsfonts}
\usepackage{algorithm}
\usepackage{algorithmic}
\usepackage{enumerate}
\usepackage{graphics}
\usepackage{subfigure}
\usepackage{amsmath}
\usepackage{courier}
\usepackage{setspace}
\usepackage{multirow}
\usepackage{booktabs}
\usepackage{soul}
\usepackage{bm}
\usepackage{threeparttable}
\usepackage{amsmath, amssymb}
\usepackage[thmmarks, amsmath, thref]{ntheorem}

\usepackage{color,xcolor}

\newtheorem{prop}{Proposition}

\usepackage{listings}

\definecolor{DarkGreen}{rgb}{0.0,0.4,0.0}

\newcommand{\cmtt}{\fontfamily{cmtt}\selectfont}

\journal{Software Impacts}

\begin{document}

\begin{frontmatter}



\title{\textbf{svds-C}: A Multi-Thread C Code for Computing Truncated Singular Value Decomposition}


\author{Xu Feng}
\author{Wenjian Yu\corref{author}}
\author{Yuyang Xie}

\cortext[author] {Corresponding author. This work was supported by NSFC (No. 61872206).\\\textit{E-mail address:} yu-wj@tsinghua.edu.cn (Wenjian Yu)}
\address{Dept. Computer Science and Technology, BNRist, Tsinghua University, Beijing, China}

\begin{abstract}

This article presents \textbf{svds-C}, an open-source and high-performance C program for accurately and robustly computing truncated SVD, e.g. computing several largest singular values and corresponding singular vectors. We have re-implemented the algorithm of \texttt{svds} in Matlab in C based on MKL or OpenBLAS and multi-thread computing to obtain the parallel program named \textbf{svds-C}. \textbf{svds-C} running on shared-memory computer consumes less time and memory than \texttt{svds} thanks to careful implementation of multi-thread parallelization and memory management. Numerical experiments on different test cases which are synthetically generated or directly from real-world datasets show that, \textbf{svds-C} runs remarkably faster than \texttt{svds} 
with averagely 4.7X and at most 12X speedup for 16-thread parallel computing on a computer with Intel CPU, 
while preserving same accuracy and consuming about half memory space. Experimental results also demonstrate that \textbf{svds-C} has similar advantages over \texttt{svds} on the computer with AMD CPU, and outperforms other state-of-the-art algorithms for truncated SVD on computing time and robustness.


\end{abstract}


\end{frontmatter}


	\section{Introduction}

	Singular value decomposition (SVD) plays a crucial role in data analysis and scientific computing, which is mainly used for principal component analysis (PCA) and low-rank approximation~\cite{matrix2012}.  Using truncated SVD, we can preserve the most important information of matrix data~\cite{GE201659}. 
 With the development of large-scale data analytics and physical simulation, high-performance program for accurate truncated SVD are largely demanded.
	
	Lanczos bidiagonalization process is a traditional method for computing truncated SVD~\cite{golub1965calculating}. It is derived from the Lanczos process for computing eigenvalues and eigenvectors of the symmetric matrix~\cite{matrix2012,svds}. The Lanczos process belongs to Krylov subspace method and constitutes an iterative algorithm, 
	with rapid convergence, calculating a few of largest and smallest eigenvalues and corresponding eigenvectors. Similarly, the Lanczos bidiagonalization process derives an algorithm which computes a few of the largest and smallest singular values and corresponding singular vectors~\cite{golub1965calculating}. However, the vectors computed in Lanczos process and Lanczos bidiagonalization process often suffer from the issue of losing orthogonality in floating-point arithmetic~\cite{matrix2012}. Therefore, proper re-orthogonalization scheme should be applied in practice. \texttt{svds} in Matlab is based on the Lanczos bidiagonalization process with full re-orthogonalization scheme~\cite{baglama2005augmented}, and is regarded as the standard tool for computing truncated SVD, especially for large sparse matrix. An augmented restarting scheme is also used in \texttt{svds} to reduce the memory cost while ensuring accuracy ~\cite{baglama2005augmented}. lansvd in PROPACK \cite{propack} is based on the Lanczos bidiagonalization process with partial re-orthogonalization scheme and augmented restarting, which computes truncated SVD with less amount of computation while sacrificing numerical stability. Besides, there are recent studies on improving the performance of complete or partial SVD computation on dense matrices in high performance computing circumstance~\cite{keyes2023high,sukkari2019qdwh,sukkari2016high}.   

Some widely-used or recently-developed algorithms for truncated SVD of sparse matrix are summarized in Table~1. 
lansvd is implemented in Fortran. PRIMME\_SVDS and Armadillo are two packages implemented in C/C++.
For matrix $\mathbf{A}$, PRIMME\_SVDS first uses the state-of-the-art truncated eigenvalue decomposition (EVD) library PRIMME (PReconditioned Iterative MultiMethod Eigensolver \cite{2010primme}) to compute EVD of $\mathbf{A}^\mathrm{T}\mathbf{A}$ or $\mathbf{A}\mathbf{A}^\mathrm{T}$. 
Then, the obtained results are used as input vectors to solve the truncated EVD of $\left[\begin{smallmatrix}\boldsymbol{0} & \mathbf{A}^{\mathrm{T}}\\\mathbf{A} & \boldsymbol{0}\end{smallmatrix}\right]$ using PRIMME. 
svds in Armadillo computes the truncated SVD of the sparse matrix according to the EVD of $\left[\begin{smallmatrix}\boldsymbol{0} & \mathbf{A}\\\mathbf{A}^{\mathrm{T}} & \boldsymbol{0}\end{smallmatrix}\right]$. All the programs can run parallely with multiple threads.
It should be pointed out,  the most efficient and stable algorithm of truncated singular value decomposition for large matrices has only an implementation in Matlab (i.e. \texttt{svds}). However, the parallel efficiency of Matlab program is not good. The performance of the  same algorithm implemented in C or Fortran could be much better on a multi-core computer.

        \begin{table}[h]
        \setlength{\abovecaptionskip}{0.1 cm}
        \caption{State-of-the-art truncated SVD algorithms.}
        \label{tab:table1}
        \centering
        \resizebox{\textwidth}{!}{
        \small{
                \renewcommand{\multirowsetup}{\centering}
                \begin{tabular}{ccc} 
                    \toprule
                   Algorithm & Method classification & Matrix type\\
                    \midrule
                    \texttt{svds} in Matlab  & Lanczos bidiagonalization process & general \\
                    lansvd~\cite{propack} & Lanczos bidiagonalization process  & general \\
                    PRIMME\_SVDS~\cite{wu2017primme_svds}  & Two stage SVD using PRIMME in \cite{2010primme} & general \\
                    svds in Armadillo~\cite{sanderson2016armadillo} & {general EVD} of $\left[\begin{smallmatrix}\boldsymbol{0} & \mathbf{A}\\\mathbf{A}^{\mathrm{T}} & \boldsymbol{0}\end{smallmatrix}\right]$  & sparse \\
                    \textbf{svds-C} & Re-implementation of \texttt{svds}  & general \\
                    \bottomrule 
                \end{tabular}
        }
        }
    \end{table}

   In this paper we present the functionalities of \textbf{svds-C}, a C program for computing truncated SVD accurately and robustly. \textbf{svds-C} is an open-source program to provide efficient tool for truncated SVD in a easy-to-use way. It contains the function for both sparse matrices and dense matrices, while allows default parameters or parameters defined by users as input. These features are introduced briefly in Section 2. A complete description of \textbf{svds-C} is provided in the package documentation, linked in the metadata.
   
   The rest of this paper is structured as follows. Section 2 introduces the algorithm and the efficient implementation of \textbf{svds-C}, and how to use it. In Section 3, numerical experiments are presented, which not only compare \textbf{svds-C} with \texttt{svds} on computers with different CPUs, but also compare \textbf{svds-C} with the other algorithms for truncated SVD. Finally, the impact of this packaged is addressed is Section 4, and this paper is concluded in Section 5.

   \section{Functionalities of \textbf{svds-C}}
    
    The purpose of \textbf{svds-C} is to provide an efficient program for rank-$k$ truncated SVD, i.e.
    \begin{equation}
        \mathbf{A} \approx \mathbf{A}_k =  \mathbf{U}_k\mathbf{\Sigma}_k\mathbf{V}_k^\mathrm{T},
    \end{equation}
    where $\mathbf{\Sigma}_k$ is a diagonal matrix with the largest $k$ singular values in descending order, while $\mathbf{U}_k$ and $\mathbf{V}_k$ are orthonormal matrices containing left and right singular vectors corresponding to the largest $k$ singular values. Besides, $\mathbf{A}_k$ is the best rank-$k$ approximation of $\mathbf{A}$ in both spectral and Frobenius norm~\cite{eckart1936}. We re-implement the lanczos bidiagonalization process and augmented restarting scheme in \texttt{svds} \cite{baglama2005augmented}, which is a widely-used tool for truncated SVD but with poor paralleling efficiency.

    In this section, we first briefly introduce the truncated SVD algorithm in \texttt{svds}. Then, we review the MKL~\cite{Intel} and OpenBLAS~\cite{OpenBLAS}, which are two efficient libraries of  matrix and vector computation, and present the details of \textbf{svds-C} which is derived from re-implementing \texttt{svds}. Finally, we present the interface of the \textbf{svds-C} program.
 
\subsection{{The Truncated SVD Algorithm in \texttt{svds}}}

The traditional algorithm for computing truncated SVD involves the Lanczos bidiagonalization process~\cite{golub1965calculating}, which is described as Algorithm 1.
According to the theory of matrix computation~\cite{golub1965calculating}, the singular values of $\mathbf{T}$ can well approximate the largest and smallest singular values of $\mathbf{A}$.

	
	\begin{algorithm}[h]
		\caption{Lanczos bidiagonalization process (LBP)}
		\label{alg5}
		\begin{algorithmic}[1]
			\REQUIRE  $\mathbf{A}\in\mathbb{R}^{m\times n}$, iteration parameter $t$ ($t\!\le\! n$)
			\ENSURE Orthonormal matrices $\mathbf{U}\!\in\!\mathbb{R}^{m\times t}$, $\mathbf{V}\!\in\!\mathbb{R}^{n\times t}$, bidiagonalization matrix $\mathbf{T}\!\in\!\mathbb{R}^{t\times t}$, such that $\mathbf{A}\!\approx \!\mathbf{U}\mathbf{T}\mathbf{V}^\mathrm{T}$
			\STATE $\mathbf{v}_1 = \mathrm{randn}(n, 1)$, $\mathbf{v}_1=\mathbf{v}_1/||\mathbf{v}_1||_2$, $\mathbf{T}=\mathrm{zeros}(t)$
			\STATE $\mathbf{u}_1=\mathbf{Av}_1$
			\STATE $\mathbf{T}(1, 1)=\alpha_1=||\mathbf{u}_1||_2$, $\mathbf{u}_1=\mathbf{u}_1/\alpha_1$
			\FOR {$i= 1, \cdots, t-1$}
			\STATE $\mathbf{v}_{i+1}=\mathbf{A}^\mathrm{T}\mathbf{u}_i-\alpha_i\mathbf{v}_i$
			\STATE $\mathbf{T}(i, i\!+\!1)=\beta_i=||\mathbf{v}_{i+1}||_2$, $\mathbf{v}_{i+1}=\mathbf{v}_{i+1}/\beta_i$
			\STATE $\mathbf{u}_{i+1}=\mathbf{A}\mathbf{v}_{i+1}-\beta_i\mathbf{u}_i$
			\STATE $\mathbf{T}(i\!+\!1, i\!+\!1)=\alpha_{i+1}=||\mathbf{u}_{i+1}||_2$, $\mathbf{u}_{i+1}=\mathbf{u}_{i+1}/\alpha_{i+1}$
			\ENDFOR
			\STATE $\mathbf{U}  = [\mathbf{u}_1,\cdots,\mathbf{u}_t]$, $\mathbf{V}  = [\mathbf{v}_1,\cdots,\mathbf{v}_t]$
		\end{algorithmic}
	\end{algorithm}

In floating-point arithmetic, $\mathbf{u}_{i+1}$ and $\mathbf{v}_{i+1}$ computed in Alg. 1 suffer from losing orthogonality when index $i$ increases. Therefore, a proper scheme of re-orthogonaliztion is required, e.g. full re-orthogonalization. In every iteration with full re-orthogonalization scheme, $\mathbf{u}_{i+1}$ and $\mathbf{v}_{i+1}$ are re-orthogonalized through executing $\mathbf{u}_{i+1}\!=\!(\mathbf{I}\!-\!\mathbf{U}^{(i)}\mathbf{U}^{(i)\mathrm{T}})\mathbf{u}_{i+1}$ and $\mathbf{v}_{i+1}\!=\!(\mathbf{I}\!-\!\mathbf{V}^{(i)}\mathbf{V}^{(i)\mathrm{T}})\mathbf{v}_{i+1}$, respectively, where $\mathbf{U}^{(i)}\!=\![\mathbf{u}_1,\cdots,\mathbf{u}_i]$ and $\mathbf{V}^{(i)}\!=\![\mathbf{v}_1,\cdots,\mathbf{v}_i]$.

Now, we introduce the details of truncated SVD algorithm in \texttt{svds}~\cite{baglama2005augmented}. In	\texttt{svds}, the extreme singular values and corresponding singular vectors of $\mathbf{T}$ in the  Lanczos bidiagonalization process are used to approximate those of $\mathbf{A}$. And, an  accuracy control scheme and an augmented restarting scheme are employed to ensure the accuracy. The accuracy control is based on the following Proposition~\cite{baglama2005augmented}.
    
    \begin{prop}
    Suppose \{$\mathbf{\hat{u}}_j,\mathbf{\hat{v}}_j,\hat{\sigma}_j$\} denotes the $j$-th largest singular triplet of the bidiagonal matrix $\mathbf{T}\in \mathbb{R}^{t\times t}$ obtained with Alg. 1, and \{$\mathbf{\tilde{u}}_j,\mathbf{\tilde{v}}_j,\tilde{\sigma}_j$\}, ($1\le j \le t$), is the $j$-th largest singular triplet of $\mathbf{A}$ obtained through
    \begin{equation}
        \tilde{\sigma}_j = \hat{\sigma}_j,~\mathbf{\tilde{u}}_j = \mathbf{U}\mathbf{\hat{u}}_j, ~\mathbf{\tilde{v}}_j = \mathbf{V}\mathbf{\hat{v}}_j,
    \end{equation}
    where $\mathbf{U}$ and $\mathbf{V}$ are the orthonormal matrices outputted by Alg.~1. Then, the computed  singular triplet of $\mathbf{A}$ fulfills:
    \begin{equation}
            \mathbf{A}\mathbf{\tilde{v}}_j = \tilde{\sigma}_j\mathbf{\tilde{u}}_j, ~~ ~  \mathbf{A}^{\mathrm{T}}\mathbf{\tilde{u}}_j =\tilde{\sigma}_j\mathbf{\tilde{v}}_j + \beta_t\mathbf{v}_{t+1}\mathbf{e}_{t}^{\mathrm{T}}\mathbf{\hat{u}}_j, ~(1\le j \le t)
        \end{equation}
    where $\beta_t$ and $\mathbf{v}_{t+1}$ can be computed by executing one more iteration of the loop in Alg.~1.
    $\mathbf{e}_{t}$ stands for a $t$-dimensional unit column vector in which the $t$-th element is 1.

    \end{prop}

    Proposition 1 means the approximation caused by truncating the Lanczos bidiagonalization process makes the difference between $\mathbf{A}^{\mathrm{T}}\mathbf{\tilde{u}}_j$ and $\tilde{\sigma}_j\mathbf{\tilde{v}}_j$, which is a residual of the $j$-th singular triplet of $\mathbf{A}$. This residual is used in \texttt{svds} to evaluate the accuracy of the computed singular triplets and to enable accuracy control. Proposition 1 facilitates the computation. Specifically,  the relative residual is used in \texttt{svds} and computed as:
    \begin{equation}
         \frac{{||\mathbf{A}^{\mathrm{T}}\mathbf{\tilde{u}}_j-\tilde{\sigma}_j\mathbf{\tilde{v}}_j ||_2}}{\tilde{\sigma}_j} = \frac{||\beta_t\mathbf{v}_{t+1}\mathbf{e}_t^{\mathrm{T}}\mathbf{\hat{u}}_j||_2}{\hat{\sigma}_j} = \frac{|\beta_t\mathbf{\hat{u}}_j(t)|}{\hat{\sigma}_j},
    \end{equation}
where $\hat{\sigma}_j=\tilde{\sigma}_j$ is the computed $j$-th largest singular value of $\mathbf{A}$. In this computation, only the singular value/vector of bidiagonal matrix $\mathbf{T}$ and $\beta_t$ in the Lanczos bidiagonalization process are used.

In \texttt{svds}, once the relative residual for any singular value is greater than a preset tolerance $\mathrm{tol}$ (e.g., $10^{-10}$), the augmented restarting scheme  \cite{baglama2005augmented} is applied to restart the Lanczos bidiagonalization process while preserving the first $k$ computed singular vectors \{$\mathbf{\tilde{u}}_j,\mathbf{\tilde{v}}_j$\}, $(j\le k)$ as the orthogonal basis vectors. 
According to Proposition 1, 
\begin{equation}
    \mathbf{A}[\mathbf{\tilde{v}}_1,\cdots,\mathbf{\tilde{v}}_k] = [\tilde{\sigma}_1\mathbf{\tilde{u}}_1,\cdots,\tilde{\sigma}_k\mathbf{\tilde{u}}_k].
\end{equation}
While calculating the ($k\!+\!1$)-th right orthogonal basis vector, $\mathbf{v}_{t+1}$ can be used because it is naturally orthogonal to $\mathbf{\tilde{v}}_j$, $(j\le k)$. To continue the Lanczos bidiagonalization process, the projection of  $\mathbf{Av}_{t+1}$ on the $\mathbf{\tilde{u}}_j$ $(j\le k)$ should be evaluated. 
The projection coefficient can be calculated with
\begin{equation}
    \mathbf{\tilde{u}}_j^{\mathrm{T}}\mathbf{Av}_{t+1} = \mathbf{v}_{t+1}^{\mathrm{T}}\mathbf{A}^{\mathrm{T}}\mathbf{\tilde{u}}_j=\mathbf{v}_{t+1}^{\mathrm{T}}(\tilde{\sigma}_j\mathbf{\tilde{v}}_j + \beta_t\mathbf{v}_{t+1}\mathbf{e}_{t}^{\mathrm{T}}\mathbf{\hat{u}}_j) = \beta_t\mathbf{e}_{t}^{\mathrm{T}}\mathbf{\hat{u}}_j = \beta_t\mathbf{\hat{u}}_j(t) ~,
\end{equation}
according to Proposition 1.
So, the ($k\!+\!1$)-th left orthogonal basis vector is computed by
\begin{equation}
\mathbf{u}_{k+1} = \mathbf{Av}_{t+1} - \sum_{j=1}^{k}\mathbf{\tilde{u}}_j\beta_t\mathbf{\hat{u}}_j(t).
\end{equation}
Then, 
the Lanczos bidiagonalization process continues to generate the subsequent orthogonal basis vectors and the updated matrix $\mathbf{T}$.
The truncated SVD algorithm in \texttt{svds} is described as Algorithm 2 in details.

     \begin{algorithm}[h]
    \caption{Truncated SVD in \texttt{svds}}
    \label{alg7}
    \begin{algorithmic}[1]
        \REQUIRE  $\mathbf{A}\in\mathbb{R}^{m\times n}$, rank parameter $k$, relative residual tolerance $\mathrm{tol}$, restarting parameter $r$   
        \ENSURE $\mathbf{\tilde{U}}\in\mathbb{R}^{m\times k}$, $\mathbf{\tilde{S}}\in\mathbb{R}^k$, $\mathbf{\tilde{V}}\in\mathbb{R}^{n\times k}$
        \STATE $t=\max(15, 3k)$
        \STATE $[\mathbf{U}, \mathbf{T}, \mathbf{V}] = \mathrm{LBP}(\mathbf{A}, t\!+\!1)$ \quad \quad \quad \quad \quad \quad \COMMENT{Alg. 1}
        \FOR{$i = 1, \cdots, r-1$}
        \STATE $[\mathbf{\hat{U}},\mathbf{\hat{\Sigma}},\mathbf{\hat{V}}] = \mathrm{svd}(\mathbf{T}(1\!:\!t, 1\!:\!t))$ 
        \IF {$\forall j \le k,  {|\mathbf{T}(t, t\!+\!1)\mathbf{\hat{U}}(t,j)|}/
            {\mathbf{\hat{\Sigma}}(j, j)}< \mathrm{tol}$} 
        \STATE \textbf{break}
        \ELSE
        \STATE $[\mathbf{\tilde{u}}_1,\cdots,\mathbf{\tilde{u}}_k]=\mathbf{U}(:, 1\!:\!t)\mathbf{\hat{U}}(:, 1\!:\!k)$, $[\mathbf{\tilde{v}}_1,\cdots,\mathbf{\tilde{v}}_k]=\mathbf{V}(:, 1\!:\!t)\mathbf{\hat{V}}(:, 1\!:\!k)$
        \STATE $\beta_t\!=\!\mathbf{T}(t,t\!+\!1)$,~$\mathbf{T}\!=\!\mathrm{zeros}(t\!+\!1)$,~$\mathbf{T}(1\!:\!k,1\!:\!k\!+\!1) \!=\! [\mathbf{\hat{\Sigma}}(1\!:\!k, 1\!:\!k),~(\beta_t\mathbf{\hat{U}}(t,1\!:\!k))^\mathrm{T}]$
        \STATE $\mathbf{v}_{k+1} = \mathbf{v}_{t+1}$, $\mathbf{u}_{k+1}=\mathbf{A}\mathbf{v}_{k+1}-[\mathbf{\tilde{u}}_1,\cdots,\mathbf{\tilde{u}}_k](\beta_t\mathbf{\hat{U}}(t,1\!:\!k))^\mathrm{T}$
        \STATE $\mathbf{T}(k\!+\!1, k\!+\!1) = \alpha_{k+1} = ||\mathbf{u}_{k+1}||_2$, $\mathbf{u}_{k+1} = \mathbf{u}_{k+1}/\alpha_{k+1}$
        \STATE Start the iterative loop in $\mathrm{LBP}(\mathbf{A}, t\!+\!1)$ from iteration $i=k+1$ with $\{\mathbf{\tilde{u}}_1,\!\cdots\!,\mathbf{\tilde{u}}_k,\mathbf{u}_{k+1}\}$,~$\{\mathbf{\tilde{v}}_1,\!\cdots\!,\mathbf{\tilde{v}}_k,\mathbf{v}_{k+1}\}$ and $\mathbf{T}$ to compute $\mathbf{U}, \mathbf{V}$ and $\mathbf{T}$
        \ENDIF
        \ENDFOR
        \STATE $\mathbf{S} = \mathrm{diag}(\mathbf{\hat{\Sigma}}(1\!:\!k, 1\!:\!k))$
        \STATE $\mathbf{\tilde{U}} = \mathbf{U}(:, 1\!:\!t)\mathbf{\hat{U}}(:, 1\!:\!k)$, $\mathbf{\tilde{V}}=\mathbf{V}(:, 1\!:\!t)\mathbf{\hat{V}}(:, 1\!:\!k)$
    \end{algorithmic}
    \end{algorithm}

    In Alg. 2, the augmented restarting scheme \cite{baglama2005augmented} is used in Step 8-12 which limits peak memory usage. In Step 10, $\mathbf{u}_{k+1}$ is computed based on (7). 
    Step~12 means the iterative Lanczos bidiagonalization process is restarted at the ($k\!+\!1$)-th iteration inheriting parts of obtained $\mathbf{U}$, $\mathbf{V}$ and $\mathbf{T}$ results. Besides, the re-orthogonalization scheme should be used in LBP (Alg.~1) and Step 10 to ensure the numerical stability. The parameter $r$ limits the maximum times of restarting in Alg.~2. With a suitable setting of $r$, Alg. 2 can output the singular values/vectors with good accuracy in most time. 

    According to the process of Alg.~2, truncated SVD in \texttt{svds} is compute-bound when $\mathbf{A}$ is dense for computing the matrix-vector multiplication on $\mathbf{A}$. When $\mathbf{A}$ is sparse with little nonzero elements, truncated SVD in \texttt{svds} is memory-bound for the memory-cost of $\mathbf{U}$ and $\mathbf{V}$.

\subsection{Implementation Details of \textbf{svds-C}}

	\subsubsection{Overview of MKL and OpenBLAS}
    
    MKL (Math kernel library) is supported by Intel for high-performance implementations of linear algebras~\cite{Intel}. MKL not only provides high-performance parallel implementations of functions in BLAS \cite{BLAS} and LAPACK \cite{LAPACK}, 
    but also supports several different computations such as sparse matrix-vector multiplication with CSR format. Therefore, according to those functions in MKL, we can re-implement \texttt{svds} to obtain a C program with better paralleling efficiency.
    
    Because MKL is designed specifically for Intel cores, we use OpenBLAS~\cite{OpenBLAS} as a optional library for other CPU cores such as the AMD core. OpenBLAS also provides high-performance parallel implementations of functions in BLAS and LAPACK. 
    Because OpenMP can be used in program with OpenBLAS, the sparse matrix-vector multiplication is implemented with OpenMP to achieve sufficient acceleration.
    
    \subsubsection{Efficient Implementations of \textbf{svds-C}}
	    To improve the performance of truncated SVD in \texttt{svds}, we re-implement it in C with MKL (or OpenBLAS) and multi-thread computing, which is called \textbf{svds-C} as the following pseudo code (with MKL/BLAS calls in comments). Several careful treatments are imposed in the implementation  for better performance on time and parallel efficiency with less memory usage. 


\begin{lstlisting}
function [U, S, V] = svds_C(A, k)
  t = max(15,3*k);
  [U, T, V] = LBP(A, t+1); % Alg. 1, which calls routines of cblas_dgemv and MKL_dcsrmv
  for i = 1:r-1
    [Uh, Sh, Vh] = svd(T(1:t, 1:t)); %LAPACKE_dgesvd
    tarr = abs(T(t,t+1)*Uh(t,1:k))/diag(Sh(1:k,1:k))<tol;
    if sum(tarr) == k
      break;
    else
      Uk = U(:,1:t)*Uh(:,1:k); %cblas_dgemm
      Vk = V(:,1:t)*Vh(:,1:k); %cblas_dgemm
      bt = T(t, t+1), T = zeros(t+1);
      T(1:k,1:k+1)=[Sh(1:k,1:k),(bt*Uh(t,1:k))'];
      uk = A*V(:,k+1)-Uk*(bt*Uh(t,1:k))'; %MKL_dcsrmv, cblas_dgemv
      ak = norm(uk,2), uk = uk/ak, T(k+1,k+1)=ak;
      Vk = [Vk,V(:,t+1)], Uk = [Uk,uk];
      Compute [U,T,V]=LBP(A,t+1) from i=k+1 with Uk, T, Vk;
    end if
  end for
  S = diag(Sh(1:k,1:k));
  U=U(:,1:t)*Uh(:,1:k), V=V(:,1:t)*Vh(:,1:k); %cblas_dgemm
  return U, S, V;
\end{lstlisting}

	    Firstly, we apply the following skills to reduce the peak memory usage.
	    \begin{itemize}
	        \item We mallocate the space of matrices and vectors at the beginning of \textbf{svds-C}, to avoid repeatedly mallocating and freeing space. 
	        \item We free the space of matrices and vectors immediately when they are useless.
	    \end{itemize}

	    Secondly, the following treatments are applied to improve the efficiency or make better parallelization.
	    \begin{itemize}
	        \item We use high-performance parallel implementations in MKL or OpenBLAS library as much as possible. Therefore, the matrix-matrix, matrix-vector and vector-vector operations are naturally parallelized. We use the following functions of MKL/BLAS to do the main computations in \textbf{svds-C}.
            \begin{itemize}
                \item \texttt{MKL\_dcsrmv} : Compute the sparse matrix-vector multiplication with the matrix stored in CSR format. For the version implemented with OpenBLAS, we implement sparse matrix-vector multiplication with OpenMP to gain sufficient acceleration.
                \item \texttt{cblas\_dgemv} : Compute the matrix-vector multiplication of the dense matrix, which is used for computing the re-orthogonalization in \textbf{svds-C}.
                \item \texttt{LAPACKE\_dgesvd} : Compute the singular value decomposition of a matrix, which is used to compute the SVD of the small matrix in \textbf{svds-C}.
                \item \texttt{cblas\_dgemm} : Compute the matrix-matrix multiplication of two dense matrices in \textbf{svds-C}.
            \end{itemize}
	        \item We parallelize the copy operations of matrix and vectors with OpenMP. Therefore, all copy operations in \textbf{svds-C} are well parallelized with less runtime.
	    \end{itemize}

When Alg.~2 performs with $R$ times  augmented restarting, the \emph{flop} (floating-point operation) count of Alg.~2 for the sparse matrix $\mathbf{A}$ is
    \begin{equation}
    \begin{aligned}
    &\mathrm{FC}_2 \approx  (2t+1)C_{mul}\mathrm{nnz}(\mathbf{A})+ (t^2+3t+1)C_{mul}(m+n)+
    \\
    &~~(R+1)C_{mul}(m+n)tk+
    (R+1)C_{svd}t^3+
    \\
    &~~R\{C_{mul}(\mathrm{nnz}(\mathbf{A})+mk)+(t-k)[2C_{mul}\mathrm{nnz}(\mathbf{A})+(t+k+4)C_{mul}(m+n)]\},
    \end{aligned}
    \end{equation}
    where $(2t+1)C_{mul}\mathrm{nnz}(\mathbf{A})$ and $ (t^2+3t+1)C_{mul}(m+n)$ reflects the matrix-vector multiplication and re-orthogonalization (\texttt{cblas\_dgemv}) in LBP in Line 3, $(R+1)C_{mul}(m+n)tk$ reflects the matrix-matrix multiplication (\texttt{cblas\_dgemm}) in Line 11, 12 and 21, and $(R+1)C_{svd}t^3$ reflects the SVD (\texttt{LAPACKE\_dgesvd}) in Line 5. In the braces of the right hand side of (8), $C_{mul}(\mathrm{nnz}(\mathbf{A})+mk)$ reflects the matrix-vector multiplication in Line 14, and $(t-k)[2C_{mul}\mathrm{nnz}(\mathbf{A})+(t+k+4)C_{mul}(m+n)]$ reflects the LBP with computed variables in Line 17. According to the analysis of \emph{flop} count, we can see that \texttt{cblas\_dgemm} and \texttt{LAPACKE\_dgesvd} are called for computing small matrices. So, the time cost on these two functions is small. In contrast, \texttt{MKL\_dcsrmv} and \texttt{cblas\_dgemv} are called in every iteration of LBP for matrix-vector multiplication and re-orthogonalization, which lead to large portion of time cost.

        \subsection{Interface of \textbf{svds-C}}
	
	In this subsection, we describe the most important functionality and features of \textbf{svds-C} using the C interface. Firstly, the sparse matrix should be read in as {\cmtt mat\_csr} structure, and the parameters are the following.
	\begin{itemize}
	\item {\cmtt nnz}, {\cmtt nrows} and {\cmtt ncols}: the number of nonzeros, rows and columns of the input matrix.
	\item {\cmtt values}: the array storing the value of each nonzero element in length {\cmtt nnz}.
	\item {\cmtt cols}: the array storing the column index of each nonzero element in length {\cmtt nnz}. The column index is from 1 to {\cmtt ncols}.
	\item {\cmtt pointerB} and {\cmtt pointerE}: the arrays in length {\cmtt nrows} storing the indexes of each row in {\cmtt cols}. For example, {\cmtt cols[{pointerB}[i]-1]} to {\cmtt cols[{pointerE}[i]-1]} store the column indexes of nonzero elements in $i$-th row.
	\end{itemize}
	Parameters {\cmtt cols}, {\cmtt pointerB} and {\cmtt pointerE} are designed following the requirement of the CSR matrix in MKL. Besides, to store the matrices of the results of \textbf{svds-C}, the structure {\cmtt mat} is used, and the parameters are the following.
	\begin{itemize}
	\item {\cmtt nrows} and {\cmtt ncols}: the number of rows and columns of the input matrix.
	\item {\cmtt d}: the array storing the value of matrix in size {\cmtt nrows} $\times$ {\cmtt ncols} in column-major format.
	\end{itemize}
	
	After specifying the input CSR matrix, we can call the
main functions:

     \indent\indent{\cmtt svds\_C(A, U, S, V, k)}\\
     \indent\indent{\cmtt svds\_C\_opt(A, U, S, V, k, tol, t, r)}\\
	The {\cmtt A} is the input matrix with structure {\cmtt mat\_csr}. Matrices {\cmtt U}, {\cmtt S} and {\cmtt V} are the matrices with structure {\cmtt mat} in size {\cmtt nrows} $\times$ {\cmtt k}, {\cmtt k} $\times$ 1 and {\cmtt ncols} $\times$ {\cmtt k}, and the space of them are mallocated automatically in {\cmtt svds\_C}. {\cmtt k} is the target rank of truncated SVD. In {\cmtt svds\_C\_opt}, 
	{\cmtt tol} represents the error tolerance of results, {\cmtt t} represents the dimension of subspace in Laczos bidiagonalization process, while {\cmtt r} represents the maximum of the augmented restarting. And, the three parameters can be difined by users in {\cmtt svds\_C\_opt}.
	Besides, we also give two functions for the dense input matrix:
	
	\indent\indent{\cmtt svds\_C\_dense(A, U, S, V, k)}\\
     \indent\indent{\cmtt svds\_C\_dense\_opt(A, U, S, V, k, tol, t, r)}\\
	The only difference is the input matrix {\cmtt A} is with the structure {\cmtt mat}.
	
	\section{Performance Analysis}
	
	In this section, experiments are carried out for the validation of \textbf{svds-C}.
	We compare \textbf{svds-C} with \texttt{svds}~\cite{baglama2005augmented}, lansvd~\cite{propack}, PRIMME\_SVDS~\cite{wu2017primme_svds} and svds in Armadillo~\cite{sanderson2016armadillo}. We use the codes of lansvd\footnote{\url{http://sun.stanford.edu/~rmunk/PROPACK/}} in the latest version of PROPACK in Fortran with OpenMP, the codes of PRIMME\_SVDS\footnote{\url{https://github.com/primme/primme}} in C compiled with MKL and PETSc~\cite{petsc-web-page} in OpenMPI for the best performance of PRIMME\_SVDS, and the codes of svds in Armadillo\footnote{\url{http://arma.sourceforge.net}} in C compiled with OpenBLAS and OpenMP for testing. We set the relative residual tolerance $\mathrm{tol}=10^{-10}$ for these algorithms. 
	The iteration parameter $t$ in lansvd is set $\max (15, 3k)$,  the same as that in \texttt{svds} and \textbf{svds-C}. And, the restarting parameter $r$ in \textbf{svds-C} and \texttt{svds} is set 10. The maximum of basis in PRIMME\_SVDS is $2k$ for convergence.
	
	Firstly, experiments are carried out on a Ubuntu server with two 8-core Intel Xeon CPU (at 2.10 GHz), and 512 GB RAM. The Matlab 2020b is used for \texttt{svds} in Matlab. All the programs are parallellized and we compare the runtime with single thread and 16 threads, and the peak memory usage of them. Then, we compare \texttt{svds} and \textbf{svds-C} with OpenBLAS on a Ubuntu server with two 64-core AMD EPYC 7V12 CPUs (at 3.30 GHZ), and 528 GB RAM.
    
    \subsection{Test Cases}

The test cases include matrices which are synthetically generated or directly from real-world datasets.

	
	Firstly, three types of decay patterns of the singular value are introduced in (16) to generate synthetic data, which are selected following \cite{yu2017single,yu2018efficient}. Decay1 is from \cite{Halko2011large} and Decay3 is from \cite{mary2015performance}. The singular values of Decay1 and Decay2 matrices decay asymptotically slowly, although they attenuate fast at the start, and the singular values of Decay 3 decay faster than them. Then, three  $40,000 \times 40,000$ synthetic matrices named Sparse1, Sparse2 and Sparse3 are randomly generated according to the Decay1, Decay2 and Decay3, respectively. In each matrix, there are 5 nonzero elements per row on average.
	
    \begin{equation}
    \begin{aligned}
        \mathrm{Decay1}&: \mathrm{its~singular~values~are~set~to~be} \begin{cases}
            \sigma_i=10^{-\frac{4}{19}(i-1)}, i=1,\cdots,20\\
            \sigma_i=10^{-4}/(i-20)^{0.1}, i>20\\
        \end{cases}\\
        \mathrm{Decay2}&: \mathrm{its~singular~values~are~set~to~be}~\sigma_i = i^{-2}, 1\le i\le \min{(m, n)}\\
        \mathrm{Decay3}&: \mathrm{its~singular~values~are~set~to~be}~\sigma_i = i^{-3}, 1\le i\le \min{(m, n)}\\
    \end{aligned}
    \end{equation}
    
As for the test cases directly from real-world data, the first three matrices are the same as those in \cite{pmlr-v95-feng18a}.
The first one is a large matrix from Movielens in size $270,896\times 45,115$ with 97 nonzero elements per row on average. The Aminer person-keyword matrix in size of $12,869,521 \times 323,896$ is the second one, with 16 nonzero elements per row on average. The third one is a social network matrix from SNAP~\cite{snapnets}, in size $82,168\times 82,168$ with 12 nonzero elements per row on average. 
We also consider two classical real-world multi-label matrix, i.e. Bibtex and Eurlex. Bibtex is a $7,395 \times 1,836$ matrix with 69 nonzero elements per row on average, while Eurlex is a $15,539 \times 5,000$ matrix with 237 nonzero elements per row on average. The Bibtex dataset is from a social bookmarking system while the Eurlex dataset is from documents about European Union Law.
Then, we consider the two sparse matrices from the experiments of PRIMME\_SVDS~\cite{wu2017primme_svds}.
LargeRegFile is in size $2,111,154 \times 801,374$ with 2 nonzero elements per row on average, while Rucci1 is in size $1,977,885 \times 109,900$ with 4 nonzero elements per row on average \cite{wu2017primme_svds}. They are two sparse matrices obtained from SuiteSparse matrix collection\footnote{\url{https://people.engr.tamu.edu/davis/matrices.html}}. Notice that LargeRegFile is a matrix with many repeated singular values, and Rucci1 has singular values decaying very slowly. 

Finally, we generate three $10,000\times 10,000$ dense matrices named Dense1, Dense2 and Dense3 following the singular value decay patterns Decay1, Decay2 and Decay3, respectively. The three matrices are used to validate the performance of \textbf{svds-C} compared with \texttt{svds} on the dense matrix.

In the following experiments, the rank parameter $k$ is fixed to 100 for the synthetically generated test cases. For the real-world cases, we test the algorithms with both $k=50$ and $k=100$.
	
	\subsection{Validation of \textbf{svds-C} on Computer with Intel CPU}

\begin{table}[t!]
    \setlength{\abovecaptionskip}{0.1 cm}
    \setlength{\belowcaptionskip}{0.05 cm}
    \resizebox{\textwidth}{!}{
    \begin{threeparttable}
        \caption{Comparison of \texttt{svds} and \textbf{svds-C} with single thread and 16 threads on computer with Intel CPU. Time is in unit of second and memory is in unit of GB.}
        \label{tab:table2}
        \centering
        \small{
            \begin{spacing}{0.9}
                \renewcommand{\multirowsetup}{\centering}
\begin{tabular}{@{\,}c@{\,}c@{\,}c@{\,}c@{\,}c@{\,}c@{\,}c@{\,}c@{\,}c@{\,}c@{\,}c@{\,}c@{\,}} 
                    \toprule
                    \multirow{2}{*}{Matrix} & \multicolumn{3}{c}{\texttt{svds}} & \multicolumn{3}{c}{\textbf{svds-C} (MKL)} & \multicolumn{3}{c}{\textbf{svds-C} (OpenBLAS)} & \multirow{2}{*}{SP$_1$} & \multirow{2}{*}{SP$_{16}$} \\
                    \cmidrule(lr){2-4} \cmidrule(lr){5-7} \cmidrule(lr){8-10}
                    & time$_1$ & time$_{16}$ & Mem & time$_1$ & time$_{16}$ & Mem & time$_1$ & time$_{16}$ & Mem &  \\ 
                    \midrule
                    Sparse1  & 16.7 & 6.13 & 0.43 & {11.5} & \textbf{2.14} & {0.28} & {10.7} & 3.41 & 0.26 & 1.5 & 2.9\\
                    Sparse2  & 11.5 & 4.62 & 0.41 & {6.52} & \textbf{1.24} & {0.26} & {6.74} & 1.94 &  0.26 & 1.8 & 3.7\\
                    Sparse3  & 11.5 & 3.76 & 0.43 & {6.49} & \textbf{1.33} & {0.26} & {6.58} & 1.93 & 0.26 &  1.8 & 2.8\\
                    Movielens ($k\!=\!50$)& 59.1 & 38.4 & 1.49 & {22.1} & \textbf{3.31} & {0.79} & {20.1} & 4.53 & 0.79 & 2.7 & 12\\
                    Movielens ($k\!=\!100$) & 156 & 87.4 & 1.99 & {52.9} & \textbf{8.67} & {1.26} & {53.4} & 11.6 & 1.26 & 2.9 & 10\\
                    Aminer ($k\!=\!50$) & 2866 & 1745 & 37.7 & 1014 & \textbf{208} & {22.6} & {913} & 234 & 22.3 & 2.8 & 8.4\\
                    Aminer ($k\!=\!100$) & 9708 & 4045 & 72.0 & 3080 & \textbf{739} & {42.3} & {2593} & 745 & 42.0 & 3.2 & 5.5\\
                    SNAP ($k\!=\!50$) & 11.3 & 3.76 & 0.43 & {7.24} & \textbf{1.62} & {0.28} & {6.65} & 2.39 & 0.28 & 1.6 & 2.3\\
                    SNAP ($k\!=\!100$) & 41.2 & 14.8 & 0.85 & 24.9 & \textbf{5.88} & {0.55} & {22.9} & 6.45 & 0.53 & 1.7 & 2.5\\
                    Bibtex ($k\!=\!50$) & 1.09 & 1.85 & 0.05 & 0.56 & \textbf{0.24} & {0.03} & {0.53} & 0.38 & 0.03 & 1.9 & 7.7\\
                    Bibtex ($k\!=\!100$) & 2.57 & 1.68 & 0.08 & 1.00 & \textbf{0.31} & {0.05} & {0.88} & 0.56 & 0.05 & 2.6 & 5.4\\
                    Eurlex ($k\!=\!50$) & 4.08 & 3.86 & 0.21 & {2.19} & \textbf{0.53} & {0.10} & {1.65} & 0.60 & 0.10 & 1.9 & 7.3\\
                    Eurlex ($k\!=\!100$) & 12.1 & 8.55 & 0.21 & {4.52} & \textbf{1.05} & {0.12} & {3.94} & 1.13 & 0.12 & 2.7 & 8.1 \\
                    LargeRegFile ($k\!=\!50$)
                    & 302 & 102 & 7.53 & 222 & \textbf{43.5} & {4.57} & {187} & 53.4 & 4.57 & 1.4 & 2.3\\
                    LargeRegFile ($k\!=\!100$) & 714 & 217 & 14.8 & 440 & \textbf{120} & {8.91} & {399} & 135 & 8.91 & 1.6 & 1.8\\
                    Rucci1 ($k\!=\!50$) & 353 & 145 & 5.70 & 202 & \textbf{44.4} & {3.32} & {167} & 53.5 & 3.27 & 1.7 & 3.3\\
                    Rucci1 ($k\!=\!100$) & 1192 & 360 & 11.1 & 609 & \textbf{161} & {6.45} & {520} & 172 & 6.39 & 2.0 & 2.2\\
                    Dense1  & 102 & 25.2 & 0.88 & {71.1} & \textbf{16.5} & {0.84} & {69.7} & 18.0 & 0.82 & 1.4 & 1.5\\
                    Dense2  & 71.5 & 17.9 & 0.88 & {46.0} & \textbf{10.7} & {0.82} & {40.4} & 11.7 &  0.82 & 1.6 & 1.7\\
                    Dense3  & 71.6 & 18.0 & 0.87 & {43.9} & \textbf{10.2} & {0.82} & {40.6} & 12.0 & 0.82 &  1.6 & 1.8\\
                    \midrule
                    Average & / & / & / & / & / & / & / & / & / & 2.0 & 4.7\\
                    \bottomrule 
                \end{tabular}
            \end{spacing}
        }
        \vspace{-1pt}
        \begin{tablenotes}
            \footnotesize
            \item time$_1$, time$_{16}$ and Mem denote the runtime with single thread, the runtime with 16 threads and the memory cost, respectively.
            \item[] SP$_1$ and SP$_{16}$ mean the speed-up ratios of \textbf{svds-C} with MKL to \texttt{svds} on time$_1$ and time$_{16}$, respectively.
        \end{tablenotes}
        \vspace{-4pt}
    \end{threeparttable}
    }
\end{table}

In this subsection, we compare our \textbf{svds-C} with \texttt{svds} on computer with Intel CPU. 
    Because the Matlab routines only take usage of the physical cores of CPU, i.e. 16 cores in our server, we test all these programs using 16-thread parallel computing for fair comparison. The results with single thread and 16 threads for the total 20 test cases are listed in Table 2. The smallest runtime for each case is indicated in bold.

Based on our efficient implementation, \textbf{svds-C} exhibits the average speedups of 2.0X and 4.7X over \texttt{svds} on runtime with single thread and 16 threads, respectively (see Table 2). Experiments also show that \textbf{svds-C} runs up to 3.2X and 12X faster than \texttt{svds} with single thread and 16 threads respectively. Although \textbf{svds-C} with OpenBLAS costs less runtime with single thread, \textbf{svds-C} with MKL performs better with 16 threads due to the better implementation for parallelization. Because \texttt{svds} is paralleled well when dealing with the dense matrix on computer with Intel CPU, the speed-up ratio of \texttt{svds} to \textbf{svds-C} with 16 threads is nearly the same with that with single thread.

Table 2 also shows that the peak memory of \textbf{svds-C} is less that of \texttt{svds} for all test cases. The reduction of memory on sparse matrices is nearly 50\% for all cases and up to 52.4\% for Eurlex ($k=50$). This is attributed to the memory management approach used in \textbf{svds-C}. The reduction of memory on the dense matrix is less than that on the sparse matrix, because the peak memory usage is mainly 
constituted of the memory of the input dense matrix.

\subsection{Validation of \textbf{svds-C} on Computer with AMD CPU}

In this subsection, we compare our \textbf{svds-C} with \texttt{svds} on a computer with AMD CPU. Their results with single thread and 16 threads for the total 23 test cases of the sparse matrix are listed in Table 3. Because on dense matrices \textbf{svds-C} costs less runtime with 8 threads, we record the runtime with single thread and 8 threads for the three dense matrices in Table~3. The smallest runtime for each case is indicated in bold.

         \begin{table}[t!]
         \centering
    \begin{threeparttable}
        \setlength{\abovecaptionskip}{0.1 cm}
        \caption{Comparison of \texttt{svds} and \textbf{svds-C} (OpenBLAS) with single thread and multiple threads on computer AMD CPU. Time is in unit of second and memory is in unit of GB.}
        \label{tab:table2}
        \centering
        \small{
            \begin{spacing}{0.9}
                \renewcommand{\multirowsetup}{\centering}
                \begin{tabular}{ccccccc@{\,}c@{\,}c@{\,}} 
                    \toprule
                    \multirow{2}{*}{Matrix} & \multicolumn{3}{c}{\texttt{svds}} & \multicolumn{3}{c}{\textbf{svds-C} (OpenBLAS)} & \multirow{2}{*}{SP$_1$} & \multirow{2}{*}{SP$_{16}$} \\
                    \cmidrule(lr){2-4} \cmidrule(lr){5-7}
                    & time$_1$ & time$_{16}$ & Mem & time$_1$ & time$_{16}$ & Mem &  \\ 
                    \midrule
                    Sparse1  & 9.18 & 10.2 & 0.54 & {5.46} & \textbf{2.41} & 0.27 & 1.7 & 4.2\\
                    Sparse2   & 6.37 & 8.07 & 0.51 & {2.86} & \textbf{1.29} & 0.26 & 2.2 & 6.3\\
                    Sparse3  & 6.36 & 6.94 & 0.51 & {2.98} & \textbf{1.35} & 0.26 & 2.1 & 5.1\\
                    Movielens ($k\!=\!50$) & 26.5 & 27.7 & 1.53 & {12.6} & \textbf{3.28} & 0.78 & 2.1 & 8.4\\
                    Movielens ($k\!=\!100$)& 62.2 & 61.9 & 2.62 & {31.3} & \textbf{7.27} & 1.26 & 2.0 & 8.5\\
                    Aminer ($k\!=\!50$) & 1331 & 974 & 52.1 & {456} & \textbf{120} & 22.3 & 2.9 & 8.1\\
                    Aminer ($k\!=\!100$) & 3419 & 2412 & 101 & {1182} & \textbf{348} & 42.0 & 2.9 & 6.9\\
                    SNAP ($k\!=\!50$) & 6.68 & 6.39 & 0.54 & {3.79} & \textbf{2.40} & 0.28 & 1.8 & 2.7\\
                    SNAP ($k\!=\!100$) & 18.8 & 17.0 & 1.05 & {12.6} & \textbf{6.32} & 0.53 & 1.5 & 2.7\\
                    Bibtex ($k\!=\!50$) & 1.14 & 1.05 & 0.03 & {0.35} & \textbf{0.24} & 0.04 & 3.3 & 4.4\\
                    Bibtex ($k\!=\!100$) & 2.23 & 1.81 & 0.07 & {0.58} & \textbf{0.32} & 0.05 & 3.8 & 5.7\\
                    Eurlex ($k\!=\!50$) & 3.12 & 2.21 & 0.19 & {1.32} & \textbf{0.31} & 0.10 & 2.4 & 7.1\\
                    Eurlex ($k\!=\!100$) & 6.51 & 4.38 & 0.19 & {2.96} & \textbf{0.54} & 0.12 & 2.2 & 8.1\\
                    LargeRegFile ($k\!=\!50$)& 147 & 156 & 9.84 & {111} & \textbf{32.5} & 4.57 & 1.3 & 4.8\\
                    LargeRegFile ($k\!=\!100$)& 323 & 323 & 21.3 & {184} & \textbf{76.4} & 8.92 & 1.8 & 4.2\\
                    Rucci1 ($k\!=\!50$) & 169 & 187 & 10.1 & {92.2} & \textbf{36.2} & 3.28 & 1.8 & 5.2\\
                    Rucci1 ($k\!=\!100$) & 471 & 466 & 19.9 & {266} & \textbf{94.0} & 6.39 & 3.8 & 5.7\\
                    Dense1  & 45.6 & 50.0* & 0.87 & {35.1} & \textbf{14.2}* & 0.82 & 1.3 & 3.5\\
                    Dense2   & 34.5 & 35.3* & 0.87 & {21.3} & \textbf{8.86}* & 0.82 & 1.6 & 4.0\\
                    Dense3  & 36.1 & 35.8* & 0.87 & {21.0} & \textbf{8.73}* & 0.82 & 1.7 & 4.1\\
                    \midrule
                    Average & / & / & / & / & / & / & 2.2 & 5.5\\
                    \bottomrule 
                \end{tabular}
            \end{spacing}
        }
        \vspace{-1pt}
        \begin{tablenotes}
            \footnotesize
            \item time$_1$, time$_{16}$ and Mem denote the runtime with single thread, the runtime with 16 threads and the memory cost, respectively. * represents the runtime with 8 threads.
            \item[] SP$_1$ and SP$_{16}$ mean the speed-up ratios of \textbf{svds-C} with OpenBLAS to \texttt{svds} on time$_1$ and time$_{16}$, respectively.
        \end{tablenotes}
         \vspace{-6pt}
    \end{threeparttable}
\end{table}

 Because Matlab internally uses  MKL for computation, the runtime of \texttt{svds} with multiple threads on computer with AMD CPU is more than that with single thread. The results in Table 3 show that, our \textbf{svds-C} runs up to 3.8X and 8.5X faster than \texttt{svds} with single thread and 16 threads respectively, and the average speedup ratios of \textbf{svds-C} to \texttt{svds} are 2.2X and 5.5X for the computing with single and 16 threads, respectively. 
 When setting more threads on the computer with AMD CPU, we find out that the runtime of \textbf{svds-C} decreases little or even increases for some cases. This suggests that setting 16 threads in the experiment is appropriate.
 
 From Table 3 we also see that the peak memory usage of \textbf{svds-C} is less than that of \texttt{svds} for all test cases except the smallest case Bibtex ($k=50$). The reduction of memory is nearly 50\% for most cases and up to 68.9\% (see the case Rucci1 ($k=100$)).
 
 \subsection{More Results and Comparisons}

 \begin{table}[t!]
    \centering
        \begin{threeparttable}
            \setlength{\abovecaptionskip}{0.1 cm}
            \caption{Runtimes of \textbf{svds-C} (MKL), lansvd~\cite{propack}, PRIMME\_SVDS~\cite{wu2017primme_svds} and svds in Armadillo~\cite{sanderson2016armadillo} with 16-thread parallel computing, on the computer with Intel CPU. Time is in unit of second.}
            \label{tab:table2}
            \centering
            \small{
                \begin{spacing}{0.9}
                    \renewcommand{\multirowsetup}{\centering}
                    \begin{tabular}{@{\,}c@{\,\,}c@{\,\,}c@{\,\,}c@{\,\,}c@{\,\,}c@{\,\,}}
                        \toprule
                        {Matrix} & Size ($m\times n$) & \footnotesize {\textbf{svds-C}} & \footnotesize {lansvd} & \scriptsize {PRIMME\_SVDS} & \scriptsize {Armadillo} \\ 
                        \midrule
                        Sparse1 & \multirow{3}{*}{40,000$\times$40,000} &\textbf{2.14} & 6.87 & 6.62 & 40.2\\
                        Sparse2 &  & 1.24 & 4.56 & \textbf{1.05} & 13.9\\
                        Sparse3 &  & 1.33 & 6.50 & \textbf{0.79} & 36.5\\ \midrule
                        Movielens ($k\!=\!50$) & \multirow{2}{*}{270,896$\times$45,115} & \textbf{3.31} & 32.9 & 10.0 & 121\\
                        Movielens ($k\!=\!100$) & & \textbf{8.67} & 59.0 & 22.6 & 257\\ \midrule
                        Aminer ($k\!=\!50$) & \multirow{2}{*}{12,869,521$\times$323,896} & {208} & 551 & \textbf{177} & 3888\\
                        Aminer ($k\!=\!100$) & & {739} & 1497 & \textbf{316} & 6700\\ \midrule
                        SNAP ($k\!=\!50$) & \multirow{2}{*}{ 82,168$\times$82,168} & \textbf{1.62} & 4.04 & 3.15 & 11.3\\
                        SNAP ($k\!=\!100$) &  & \textbf{5.88} & 9.85 & 11.5 & 30.2\\ \midrule
                        Bibtex ($k\!=\!50$) & \multirow{2}{*}{7,395$\times$1,836} & \textbf{0.24} & 0.55 & 0.78 & 1.90\\
                        Bibtex ($k\!=\!100$) & & \textbf{0.31} & 0.75 & 2.73 & 4.05\\ \midrule
                        Eurlex ($k\!=\!50$) & \multirow{2}{*}{15,539$\times$5,000} & \textbf{0.53} & 3.26 & 1.33 & 14.0\\
                        Eurlex ($k\!=\!100$) & & \textbf{1.05} & 6.70 & 3.53 & 23.1\\ \midrule
                        LargeRegFile ($k\!=\!50$) & \multirow{2}{*}{$2,111,154\times801,374$} & {43.5} & 53.4 & \textbf{6.00} & 170\\
                        LargeRegFile ($k\!=\!100$) & & {120} & 320 & \textbf{18.7} & 431\\ \midrule
                        Rucci1 ($k\!=\!50$) & \multirow{2}{*}{1,977,885$\times$109,900} & {44.4} & 65.7 & \textbf{19.6} & 253\\
                        Rucci1 ($k\!=\!100$) & & {161} & {158} & \textbf{50.0} & 637\\
                        \bottomrule 
                    \end{tabular}
                \end{spacing}
            }
            \vspace{-1pt}
        \end{threeparttable}
    \end{table}
    
    In this subsection, we compare our \textbf{svds-C} program with other programs for truncated SVD, i.e. 
    lansvd, PRIMME\_SVDS and svds in Armadillo. 
   We test all these programs using 16-thread parallel computing for fair comparison on the computer with Intel CPU. The runtime for the total 23 test cases of sparse matrices is listed in Table 4. The smallest runtime for each case is indicated in bold. The computed singular values are plotted in Fig. 1.

     \begin{figure}[t!]
     \centering
     \subfigure[Sparse1] 
    {
        \includegraphics[width=4.2cm,trim=93 265 113 275,clip]{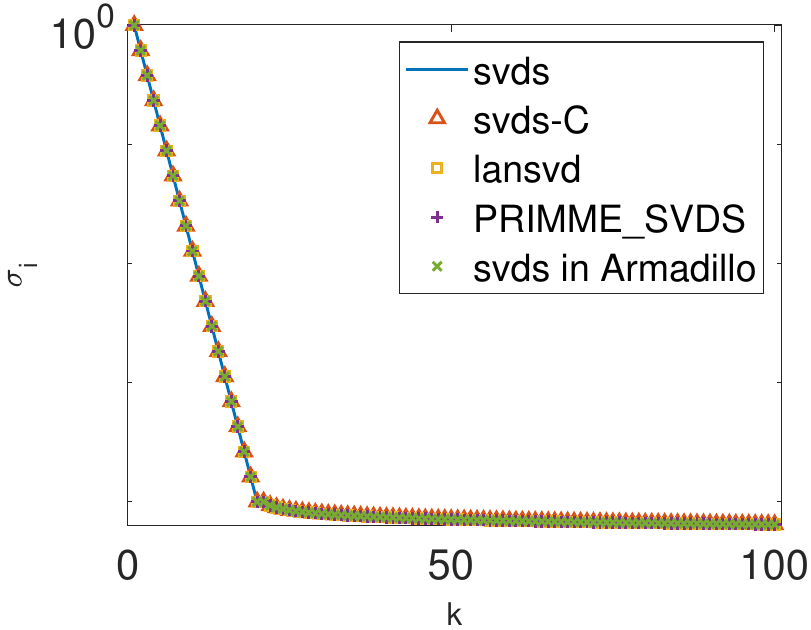} } 
    \subfigure[Sparse2] {
        \includegraphics[width=4.2cm,trim=93 265 113 275,clip]{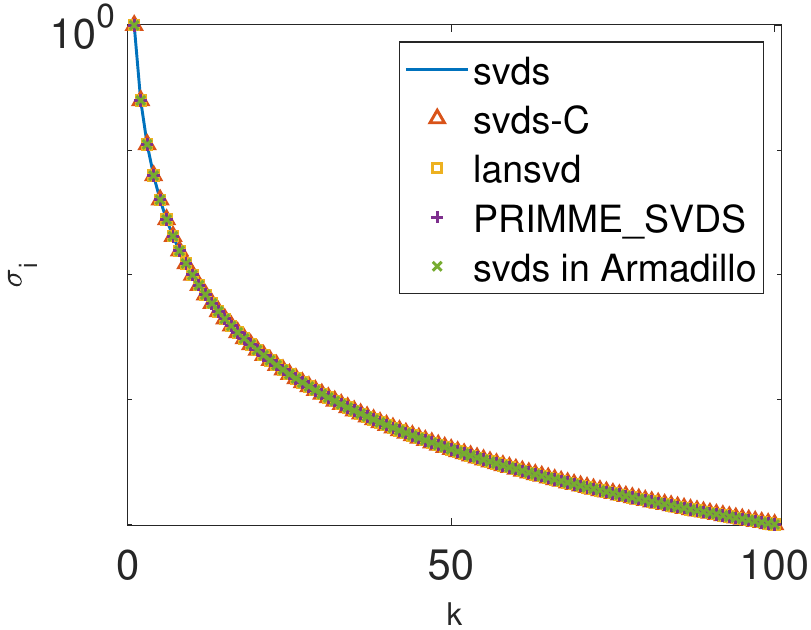} 
    }
    \subfigure[Sparse3] {
        \includegraphics[width=4.2cm,trim=93 265 113 275,clip]{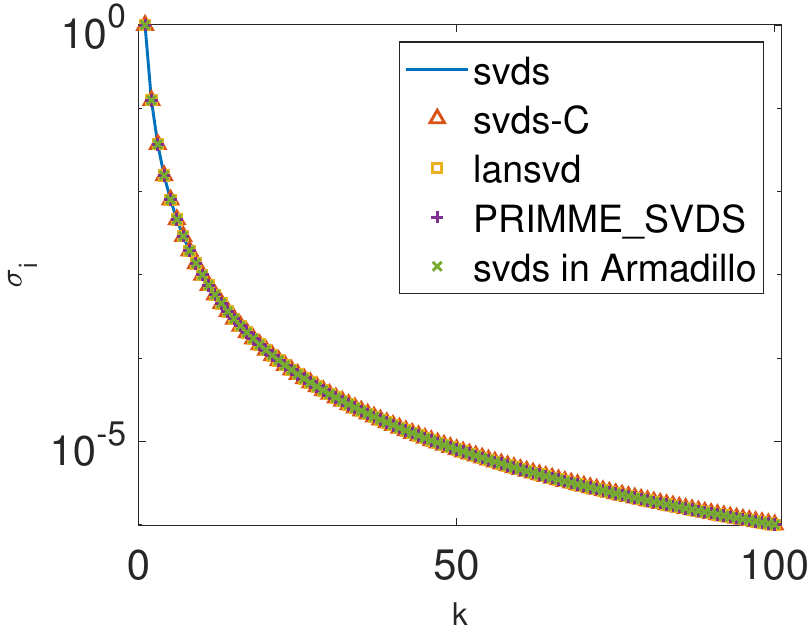} 
    }
    \subfigure[Movielens ($k=100$)] {
        \includegraphics[width=4.2cm,trim=93 265 113 275,clip]{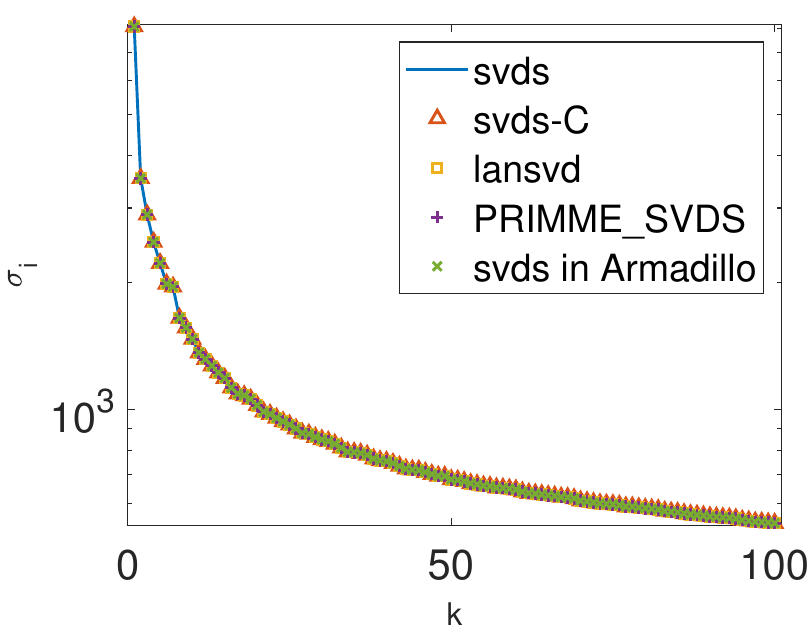} 
    }
    \subfigure[Aminer ($k=100$)] {
        \includegraphics[width=4.2cm,trim=93 265 113 275,clip]{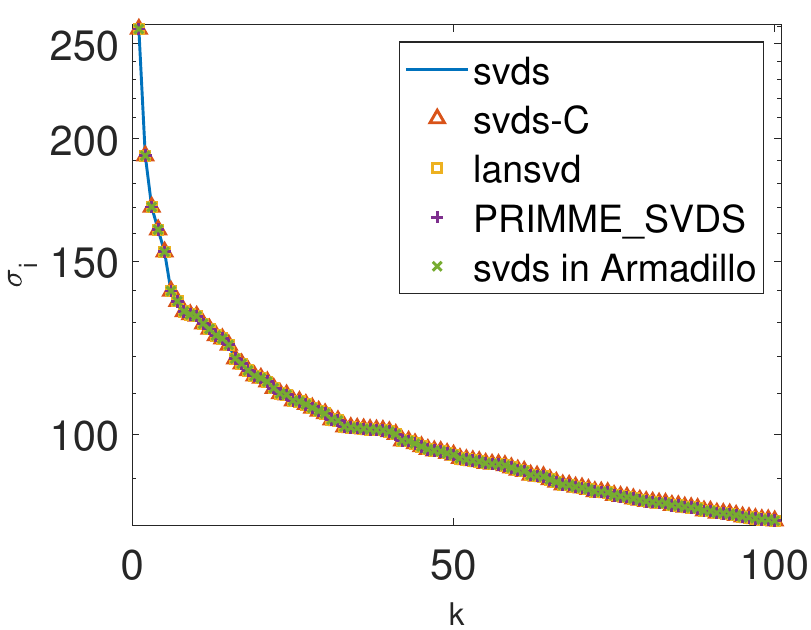} 
    }
    \subfigure[SNAP ($k=100$)] { 
        \includegraphics[width=4.2cm,trim=93 265 113 275,clip]{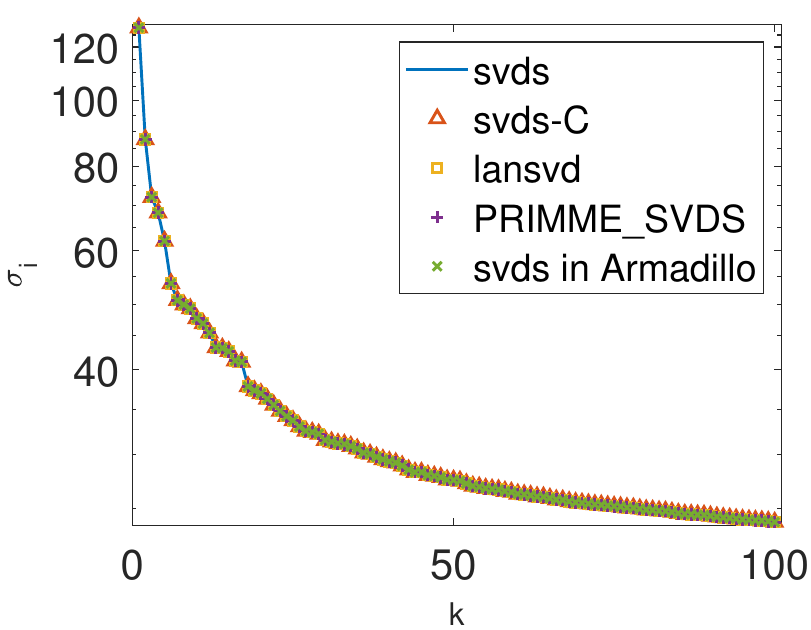} 
    }
    \subfigure[Bibtex ($k=100$)] {
        \includegraphics[width=4.2cm,trim=93 265 113 275,clip]{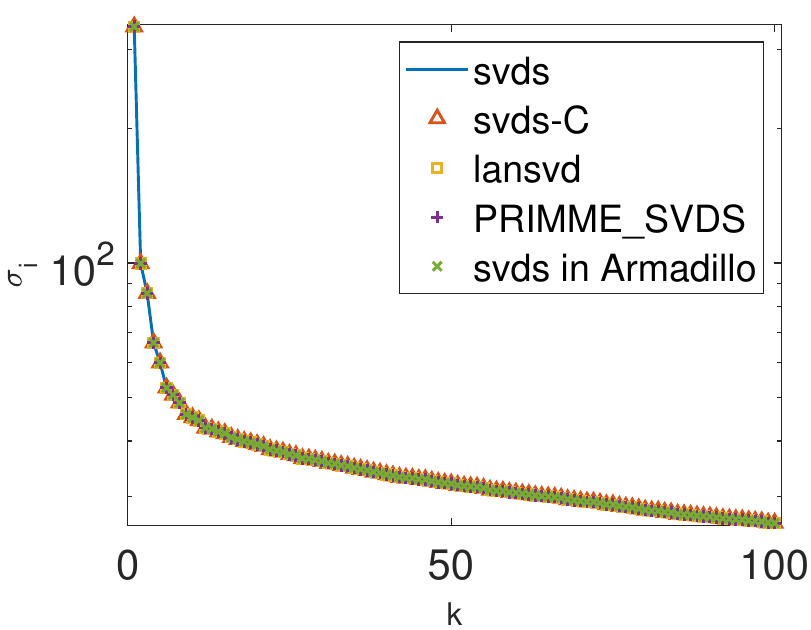} 
    }
    \subfigure[Eurlex ($k=100$)] {
        \includegraphics[width=4.2cm,trim=93 265 113 275,clip]{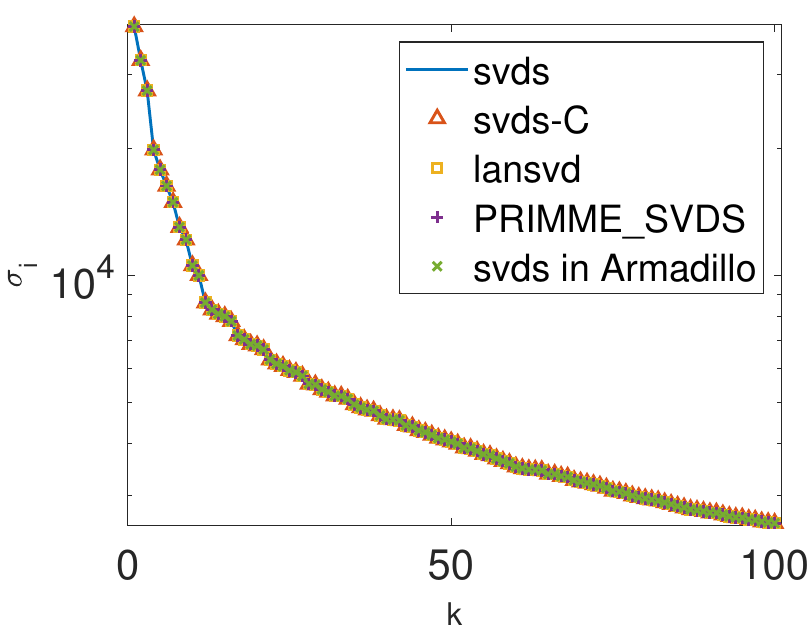} 
    }
    \subfigure[LargeRegFile ($k=100$)] { 
        \includegraphics[width=4.2cm,trim=93 265 113 275,clip]{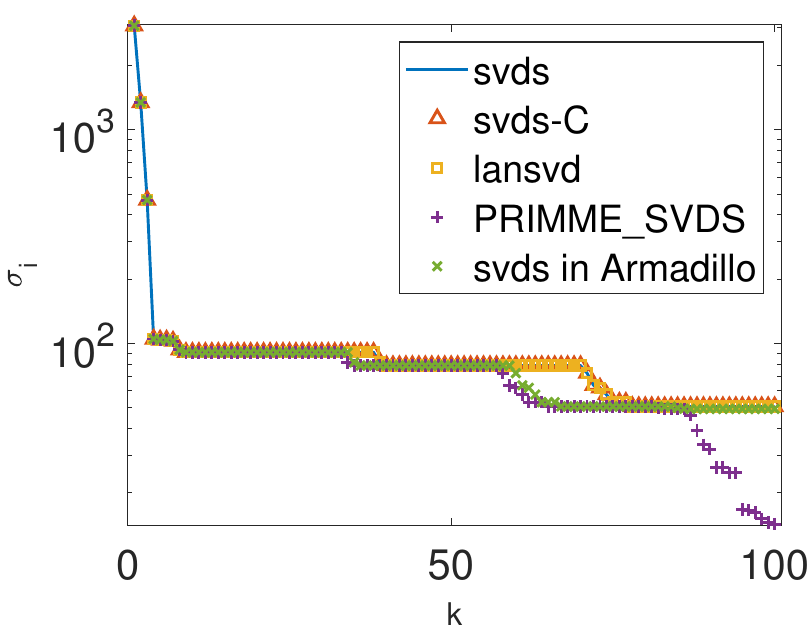} 
    }
    \subfigure[Rucci1 ($k=100$)] {
        \includegraphics[width=4.2cm,trim=93 265 113 275,clip]{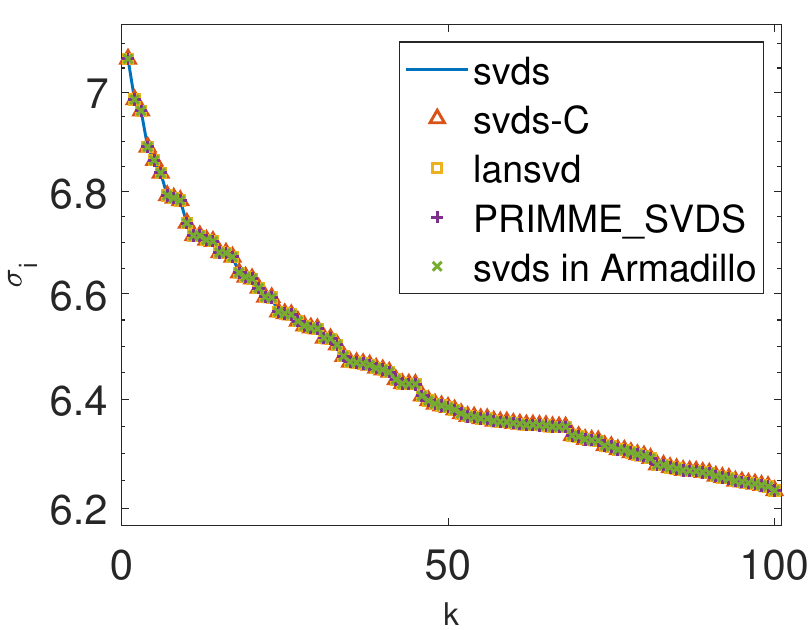} 
    }
    \caption{The computed singular values of test cases from \texttt{svds}, \textbf{svds-C}, lansvd~\cite{propack}, PRIMME\_SVDS~\cite{wu2017primme_svds} and svds in Armadillo~\cite{sanderson2016armadillo} (setting $k=100$).}
\end{figure} 

  From Table~4 we see that the runtime with 16 threads of \textbf{svds-C} is the smallest or close to the smallest for most cases. For the cases with $m \gg n$ (Aminer, LargeRegFile and Rucci1), PRIMME\_SVDS  costs less runtime than \textbf{svds-C} because it computes the eigenvalue decomposition of $\mathbf{A}^\mathrm{T}\mathbf{A}$ at the first stage. From Fig. 1, we see that the results of  \textbf{svds-C}  completely matches \texttt{svds} which validates the accuracy of \textbf{svds-C}. However, lansvd, PRIMME\_SVDS and svds in Armadillo all induce accuracy issue on the case LargeRegFile which has many multiple singular values. As shown in Fig. 1, their results are not correct, largely different from those of \texttt{svds}. Therefore, \textbf{svds-C} is an efficient choice for calculating accurate truncated SVD robustly.

	\section{Impact}

    Truncated SVD, e.g. computing several largest singular values and corresponding singular vectors, is a very important factorization in matrix computations, which has been used in the area of machine learning and data mining for solving problems of  dimension reduction, information retrieval and feature selection, etc. For example, it is used in NetMF  to compute the network embedding~\cite{qiu2018network}, and is repeatedly applied in the singular value thresholding (SVT) algorithm for matrix completion \cite{Cai2010}. Among the algorithms for truncated SVD, \texttt{svds} is the most widely-used and most robust tool. However, \texttt{svds} does not perform well on multi-thread computing because it is implemented in Matlab. And, there is an incredible shortcoming that there is no \texttt{svds} algorithm implemented in C/C++. Our work (\textbf{svds-C}) fills this gap.
    The  experimental results in this paper show the poor paralleling efficiency of \texttt{svds}, and that \textbf{svds-C} runs remarkably faster than \texttt{svds} under the setting of parallel computing on a computer with no matter Intel CPU or AMD CPU. Besides, we provide the source code and building instructions of the \textbf{svds-C} to the research community through a public release on GitHub, which enables researchers from various areas to  efficiently compute truncated SVD in their customized and real applications.
    

    \section{Conclusion}
    
    This article presents the open-source program \textbf{svds-C}, a high-performance C program that provides functionalities to compute truncated SVD accurately and robustly. We re-implement the truncated SVD algorithm \texttt{svds}in Matlab to obtain a parallel C program named \textbf{svds-C} based on the MKL~\cite{Intel} and OpenBLAS~\cite{OpenBLAS}. Numerical experiments on synthetically generated or real-world test matrices
    show that \textbf{svds-C} runs up to 3.2X and 12X faster than \texttt{svds} on the machine with Intel CPU for single-thread and 16-thread computing respectively, while remarkably reducing the memory cost. On the computer with AMD CPU, \textbf{svds-C} runs up to 3.8X and 8.5X faster than \texttt{svds} for single-thread and 16-thread computing,  respectively. Our experiments also validates that \textbf{svds-C} outperforms the other algorithms for truncated SVD on computing time and robustness. As the result,  \textbf{svds-C} can be regarded as a good replacer of \texttt{svds} for computing truncated SVD of large-scale matrices.

\bibliographystyle{elsarticle-num}
\bibliography{fastsvdpack}



\section*{Current code version}
\label{}


\begin{table}[!h]
\begin{tabular}{|l|p{6.5cm}|p{6.5cm}|}
\hline
\textbf{Nr.} & \textbf{Code metadata description} & \textbf{Please fill in this column} \\
\hline
C1 & Current code version & v1 \\
\hline
C2 & Permanent link to code/repository used for this code version & \url{https://github.com/THU-numbda/svds-C} \\
\hline
C3  & Permanent link to Reproducible Capsule & \\
\hline
C4 & Legal Code License   & MIT \\
\hline
C5 & Code versioning system used & git \\
\hline
C6 & Software code languages, tools, and services used & C \\
\hline
C7 & Compilation requirements, operating environments \& dependencies & MKL in oneAPI~\cite{Intel} or OpenBLAS~\cite{OpenBLAS}\\
\hline
C8 & If available Link to developer documentation/manual & \url{https://github.com/THU-numbda/svds-C}\\
\hline
C9 & Support email for questions & \\
\hline
\end{tabular}
\caption{Code metadata (mandatory)}
\label{} 
\end{table}




\end{document}